\documentclass[aps,prl,twocolumn,reprint,superscriptaddress]{revtex4-2}
\usepackage{graphicx}
\usepackage{amsmath}
\usepackage{amssymb}
\usepackage{colordvi}
\usepackage{mathrsfs}
\usepackage{bm}
\usepackage{verbatim}
\usepackage{dcolumn}
\usepackage{epsfig}
\usepackage{subfigure}
\usepackage{appendix}
\usepackage[colorlinks,allcolors=blue]{hyperref}

\begin{document}
\title{Altermagnetism Induced Topological Phase Transitions in Kane-Mele Model}
\author{Zhengtian Li}
\affiliation{International Center for Quantum Design of Functional Materials,
CAS Key Laboratory of Strongly-Coupled Quantum Matter Physics, and Department of Physics,
University of Science and Technology of China, Hefei, Anhui 230026, China}
\affiliation{Hefei National Laboratory, University of Science and Technology of China, Hefei 230088, China}
\author{Zeyu Li}
\email[Correspondence author:~]{lzy92@ustc.edu.cn}
\affiliation{International Center for Quantum Design of Functional Materials,
CAS Key Laboratory of Strongly-Coupled Quantum Matter Physics, and Department of Physics,
University of Science and Technology of China, Hefei, Anhui 230026, China}
\affiliation{Hefei National Laboratory, University of Science and Technology of China, Hefei 230088, China}
\author{Zhenhua Qiao}
\email[Correspondence author:~]{qiao@ustc.edu.cn}
\affiliation{International Center for Quantum Design of Functional Materials,
CAS Key Laboratory of Strongly-Coupled Quantum Matter Physics, and Department of Physics,
University of Science and Technology of China, Hefei, Anhui 230026, China}
\affiliation{Hefei National Laboratory, University of Science and Technology of China, Hefei 230088, China}
\date{\today}
	
\begin{abstract}
We theoretically demonstrate that Chern number tunable quantum anomalous Hall effect (QAHE) and second-order topological insulators can be induced in the two-dimensional $\mathbb{Z}_2$ topological insulator (TI), i.e., Kane-Mele model, by applying $d$-wave altermagnetism. When the N\'eel vector of altermagentism lies in the $x-y$ plane, the $\mathbb{Z}_2$ TI is broken and driven into a second-order topological insulator phase, exhibiting the representative corner states at nanoflakes. When the intrinsic Rashba spin-orbit coupling is further included, the second-order TI is further driven into the QAHE phase with various Chern numbers (e.g., $\mathcal{C}=\pm1$ or $\pm3$). When the N\'eel vector is along $z$ direction, the intrinsic Rashba spin-orbit coupling is necessary to break the mirror symmetry to allow a sequential emergence of second-order TI and QAHE along with the increase of altermagentism strength. We also observe the QAHE with mixed-chirality, i.e., there exist counter-propagating edge modes but net chiral current at the ribbon boundary. Our work shows that altermagnetism can play a crucial role in exploring a rich variety of topological phases, just like its counterparts of ferromagnetism and antiferromagnetism.
\end{abstract}

\maketitle

\textit{Introduction}.---Recently, some theoretical works unveiled the existence of a new kind of collinear magnetism named as altermagnetism, which is characterized by spin-polarized band structures with compensated antiferromagnetic configuration~\cite{Smejkal2022conventional, Smejkal2022emerging}. The basic properties of altermagnetism can be captured by spin symmetries~\cite{Liu2022Spin,Chen2024Enumeration}. By using the first-principles calculations, several materials have been predicted as altermagnets, such as $\rm RuO_2$~\cite{Smejkal2020crystal}, $\rm MnF_2$~\cite{Yuan2020giant}, $\rm Mn_5Si_3$~\cite{Reichlova2021macroscopic}, and $\rm La_2CuO_4$~\cite{lane2018anti}. Some experiments have further identified the signatures of altermagnetism, i.e., non-relativistic spin splitting in $\rm MnTe_2$~\cite{Zhu2024observation} and $\rm MnTe$~\cite{Krempasky2024altermagnetic}. In these materials, nonmagnetic atoms play a crucial role in breaking the translational or inversion symmetry, resulting in spin-splitting in both real and reciprocal spaces in contrast to those of antiferromagnets. Due to its superior properties, altermagnetic materials are considered to be ideal platforms to explore novel physical phenomena such as Majorana zero modes~\cite{Say2024Alter, Zhu2023topological, Li2023majorana}, topological matter states~\cite{Li2024creationa, Ezawa2024detecting, Parshukov2024topological}, Andreev reflection~\cite{Sun2023andreeva, Papaj2023andreeva}, and magnetic tunnel junctions~\cite{Chi2024crystalfacetoriented}. 

Magnetism has been extensively used to design various topological states, because it can break the time-reversal
symmetry ($\mathcal{T}$). For example, by introducing the out-of-plane ferromagnetism in two-dimensional $Z_2$ topological insulators (TI) and graphene systems via doping magnetic atoms or proximity-coupled with magnetic substrates~\cite{Haldane1988, Ren2016topological,Yu2010quantized,Qiao2010quantum,Kevin2013Chern,Quantum2015Jia,Weng2015Quantum,Liu2016Quantum,Chang2013Experimental,Checkelsky2014Trajectory,Kou2014Scale,Chang2015High,Pan2014valleypolarized,Tse2011Quantum,Qiao2012,Ezawa2013,Ezawa2012valleypolarized,Liu2023,Shen2024}, the quantum anomalous Hall effect (QAHE) can be realized; By applying the in-plane ferromagnetism in two-dimensional $Z_2$ TI~\cite{Ren2020engineering,Chen2020Uni,LiuZheng}, the second-order TI (i.e., corner states) can be engineered; In antiferromagnetic materials, such as intrinsic antiferromagnetic topological insulator $\rm MnBi_2Te_4$~\cite{otrokov2019prediction,otrokov2019unique,yu2020quantum,dong2019topological,chang2020robust,jiang2019intrinsic,shi2020berry,hekecpl,Liang2024Chern,Lizeyu2025} and axion insulator EuIn$_2$As$_2$~\cite{Riberolles2021Magetic,yuan2019higher,yang2020inplane}, the QAHE can also be obtained. Therefore, a natural question arises: as the third collinear magnetism, whether the altermagnetism can also be used to produce some topological states?

In this Letter, we theoretically demonstrate that the introduction of $d$-wave altermagnetism into the  $\mathbb{Z}_2$ TI, i.e., Kane-Mele model, can induce various topological phases, e.g., second-order TI and QAHE with various Chern numbers. When the N\'eel vector of altermagentism lies in $x-y$ plane, the system is driven into the second-order TI phase from $\mathbb{Z}_2$ TI. When the intrinsic Rashba spin-orbit coupling is further considered, the system gradually enters the QAHE with Chern numbers of $\mathcal{C}=\pm 1$ or $\mathcal{C}=\pm 3$, depending on the altermagnetism. When the N\'eel vector of altermagentism is along $z$ direction, the system is sequentially driven into three kinds of topological phases along with the increase of altermagnetism, where intrinsic Rashba spin-orbit coupling is necessary to break the mirror symmetry. It is noteworthy that, in a unconventional QAHE phase, there are three pairs of edge states with mixed chiralities, resulting in $\mathcal{C}=\pm 1$. These findings provide a versatile platform for studying a rich variety of topological phases induced by altermagnetism in Kane-Mele model. 

\textit{Model Hamiltonian and Symmetry Analysis}.---The model Hamiltonian of a modified Kane-Mele model in the presence of $d$-wave altermagnetism can be written as $H = H_{\rm KM} + H_{\rm AM}$, with
\begin{eqnarray}
		H_{\rm KM}&=& -t \sum_{\langle i j\rangle} c_i^{\dagger} c_j+i t_{\mathrm{SO}} \sum_{\langle\langle i j\rangle\rangle} \nu_{i j} c_i^{\dagger} s_z c_j \nonumber 
		 \\
		 & +& i t_{\mathrm{IR}}\mu \sum_{\langle \langle i j\rangle \rangle  \alpha \beta} \hat{\mathbf{e}}_z \cdot\left(\boldsymbol{s}_{\boldsymbol{\alpha} \boldsymbol{\beta}} \times \boldsymbol{d}_{i j}\right) c_{i \alpha}^{\dagger} c_{j \beta}, \label{equ-1} \\
     H_{\rm AM} &=& \lambda \sum_{\langle\langle i j  \rangle \rangle} c_i^{\dagger}  \Delta \boldsymbol{s} \cdot \boldsymbol{\hat n} c_j, \label{equ-2}
\end{eqnarray} 
where $\Delta = \cos(2\theta)$, $c_i^{\dagger} = (c_{i\uparrow}^{\dagger},c_{i\downarrow}^{\dagger})$ is the creation operator for an electron with spin-up/down ($\uparrow / \downarrow$) at $i$-th site. $v_{i j}=\boldsymbol{d}_j \times \boldsymbol{d}_i /\left|\boldsymbol{d}_j \times \boldsymbol{d}_i\right|$, where $\boldsymbol{d}_i$ and $\boldsymbol{d}_j$ are the bonds between two next-nearest neighbor sites. In Eq.~(\ref{equ-1}), the first two terms are the nearest-neighbor hopping with an amplitude of $t$ and intrinsic spin-orbit coupling, respectively, giving rise to the seminal $\mathbb{Z}_2$ TI~\cite{Kane2005topological}. The third term is intrinsic Rashba spin-orbit coupling in low-buckled honeycomb lattice~\cite{Ezawa2012valleypolarized, Pan2014valleypolarized, Ren2016quantum}. Equation~(\ref{equ-2}) describes the $d$-wave altermagnetism term $(k_x^2-k_y^2)\boldsymbol{s} \cdot \boldsymbol{\hat n}$ introduced by the substrate.
To ensure the $d$-wave symmetry, we adopt the anisotropic hopping between same sublattices, which resembles the form of d-wave superconductivity in honeycomb lattice~\cite{AEdge,AChiral,sm}. $\boldsymbol{s}$ are the Pauli matrices of spin degree of freedom. $\boldsymbol{\hat n}$ represents the direction of N\'eel vector~\cite{Smejkal2022emerging}. $\theta$ is the azimuthal angle of $\boldsymbol{d}_{i j}$, which is a unit vector pointing from $j$-th to $i$-th site. 

\begin{figure}[t]
\centering
\includegraphics[width=8.0 cm,angle=0]{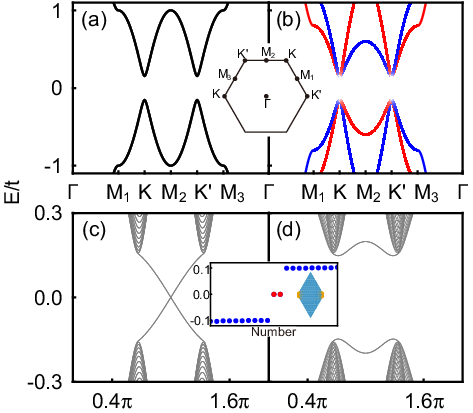}
\caption{(a-b) Bulk band structures (a) without and (b) with in-plane altermagnetism along $y$ direction. After introducing in-plane altermagnetism, the band structure is split into spin-down and -up channels labeled by blue and red lines, respectively. Inset: The first Brillouin zone with high-symmetry points. (c-d) Corresponding zigzag-edged ribbon band structures (c) without and (d) with in-plane altermagnetism. Inset: Energy spectra and local density of states of in-gap corner states. Other parameters are set to be $\lambda=0.1t$, and $t_{\rm SO}=0.03t$.}
\label{Fig1}
\end{figure}

Let us first analyze the system symmetry before further discussing the topological states induced by altermagnetism. In our preview work~\cite{Ren2016quantum}, we demonstrated that QAHE can survive in a honeycomb lattice when the system simultaneously breaks the joint symmetries of $\mathcal{T} \otimes \mathcal{M}_z$ and $\mathcal{T} \otimes \mathcal{M}_z \otimes \mathcal{I}$, where $\mathcal{M}_z$ and $\mathcal{I}$ represent the out-of-plane mirror reflection and inversion symmetries, respectively. If the N\'eel vector of altermagnetism lies in $x-y$ plane, $H_{\rm AM}$ has even parity under the operation of $\mathcal{T} \otimes \mathcal{M}_z \otimes \mathcal{I}$, resulting in a vanishing Chern number. To obtain a non-vanishing Chern number, the intrinsic Rashba spin-orbit coupling, possessing odd parity under $\mathcal M_z$ and even parity under $\mathcal{I}$, is necessary. In contrast, if the N\'eel vector of altermagnetism has a $z$ component, the joint symmetries of $\mathcal{T} \otimes \mathcal{M}_z$ and $\mathcal{T} \otimes \mathcal{M}_z \otimes \mathcal{I}$ are simultaneously broken.

\begin{figure*}[!htp]
\centering
\includegraphics[width=17 cm,angle=0]{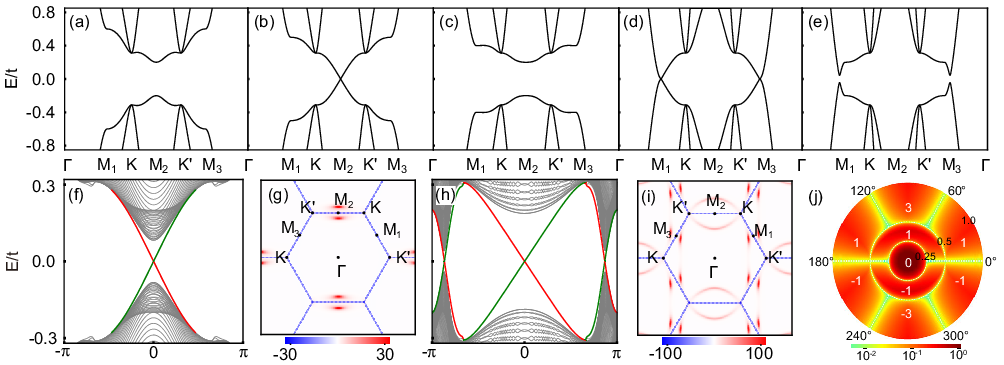}
\caption{Evolution of bulk band structures of Kane-Mele model as a function of altermagnetism at fixed intrinsic spin-orbit coupling $t_{\rm SO}=0.06t$ and intrinsic Rashba spin-orbit coupling $t_{\rm IR}=0.06t$ with N\'eel vector along $y$ direction. (a)-(e) $\lambda/t = 0.2, 0.25, 0.3, 0.5, 0.6$, respectively. (f) and (h) Band structures of zigzag nanoribbons with altermagentism strength $\lambda/t = 0.3, 0.6$, respectively. Edge states located at opposite boundaries are highlighted by red and green curves, respectively. (g) and (h) Corresponding Berry curvatures distribution for $\lambda = 0.3t, 0.6t$, respectively. (j) Phase diagram as a function of the azimuthal angle $\phi$ of the N\'eel vector of altermagnetism with $t_{\mathrm{IR}} = t_{\mathrm{SO}} = 0.10t$. Color bar measures bulk band gap amplitude.}
\label{Fig2}
\end{figure*}

\textit{Second-Order Topological Insulator}.--- Figure~\ref{Fig1} displays the bulk band structures of the Kane-Mele model system with and without the in-plane altermagnetism. When only the intrinsic spin-orbit coupling is turned on, the system opens a non-trivial band gap to host the $\mathbb{Z}_2$ TI with double spin degeneracy. When the in-plane altermagnetism is further considered, the spin-polarized band structure splits alternatively along the high-symmetry lines, signifying the characteristics of altermagnet. When the altermagnetism is rotated to $z$ direction, the bulk band structure also splits alternatively along the high-symmetry lines~[see Fig.~S1(b) in Supplemental Materials~\cite{sm}]. During the operating process, the bulk band gap retains open, usually implying no topological phase transition. However, due to $\mathcal{T}$ breaking, one can find that the edge states open a sizeable band gap, which is reminiscent of topological phase transition from first- to second-order TI. To confirm its topological properties, one can calculate the mirror-graded Zak phase. For simplicity, we set the N\'eel vector of in-plane altermagnetism along $y$ direction. The system exhibits mirror symmetry $\mathcal{M}_y = s_xi\sigma_y$~\cite{dresselhaus2007group, xinzheng2019group}, commuting with bulk Hamiltonian with a vanishing lattice momentum in the $y$ direction, i.e., $[\mathcal{M}_y, H(k_x,0)]=0$. Therefore, one can divide the Hamiltonian by using a unitary transformation $\mathcal{U}$, which is the eigenvector of operation $\mathcal{M}_y$ written as $\mathcal{U}^\dagger H(k_x,0) \mathcal{U} = H_+(k_x)\oplus H_-(k_x)$,
and Zak phase can be obtained by
\begin{equation}
	\gamma=i\sum_n\int_{BZ} \langle nk|\partial_k |n k\rangle,
\end{equation}
where $|n k\rangle$ is the eigenvector and $n$ is the index of occupied states. By using the Wilson loop method~\cite{Sheng2019twodimensional, Zak1989berry}, we obtain two Zak phases $\gamma_{\pm}=\pm \pi$, signaling that the system is a second-order TI. Moreover, in the presence of intrinsic Rashba spin-orbit coupling, the out-of-plane altermagnetism can also open an edge state gap possessing the second-order TI [See Figs.~S1(c) and S1(d)]. 

For in-plane altermagnetism, two in-gap corner states are distributed at the obtuse angles displayed in orange [see Inset of Figs.~\ref{Fig1}(c) and \ref{Fig1}(d)]. For out-of-plane altermagnetism, two corner states are formed at acute angles as displayed in Fig.~S1(e). However, the two in-gap corner states have an energy shift due to the lack of chiral symmetry. Our symmetry analysis indicates that the corner states for in-plane and out-of-plane cases are respectively protected by mirror-reflection symmetry $\mathcal{M}_y$ and joint symmetry of $\mathcal{M}_x$ and $\mathcal{T}$.

\textit{Quantum Anomalous Hall effect}.---With further increase of in-plane altermagnetism, the bulk band structure evolves dramatically, i.e., the band gap closes and reopens, implying that topological phase transition occurs. As displayed in Figs.~\ref{Fig2}(a)-\ref{Fig2}(e), the bulk band gap closes and reopens twice at the critical points of $\lambda_{\rm C}=0.25t$, and $0.50t$, respectively. The first band gap closing occurs only at M$_2$ point, while the second one occurs at both M$_1$ and M$_3$ points. Figures~\ref{Fig2}(f) and \ref{Fig2}(h) display the corresponding zigzag-nanoribbon band structures. After the first band gap reopening, a pair of chiral edge states appears. Based on the method of Wilson loop~\cite{ChernNumber}, the calculated Chern number $\mathcal{C} = 1$ confirms that the system enters the QAHE phase. Similarly, the system enters another QAHE phase with $\mathcal{C} = 3$ exhibiting three pairs of chiral edge modes. 
To shed light on these topological phase transitions, a phase diagram in the ($t_{\mathrm{IR}}$, $\lambda$) plane is drawn for altermagnetism along $y$ direction~[See Fig.~S4(a)]. Figures~\ref{Fig2}(g) and \ref{Fig2}(i) plot the Berry curvature distribution of the QAHE phases with $\mathcal{C} = 1$ and $\mathcal{C} = 3$, respectively. In Fig.~\ref{Fig2}(g), the Berry curvatures are mainly located around $\rm M_2$ points instead of $\rm K$ and $\rm K'$ points of the Kane-Mele model; While in Fig.~\ref{Fig2}(i), the Berry curvatures with the same sign are located around all three different $\rm M$ points.

Figure~\ref{Fig2}(j) draws the phase diagram of Chern number $\mathcal{C}$ as a function of the azimuthal angle $\phi$ of the N\'eel vector and strength $\lambda$ of altermagnetism at fixed t$_{\mathrm{IR}}$ and t$_{\mathrm{SO}}$, where the color measures the bulk band gap amplitude. When $0<\lambda<0.25t$, the second-order TI forms a circular region. When $\lambda>0.25t$, the system enters the QAHE phase. 
It is striking that the first and second QAHE ($\lambda>0.25t$) phases are divided into two ($\lambda<0.5t$) and six ($\lambda>0.5t$) sections, respectively, which differs from the previous works~\cite{Ren2016quantum,Li2022Chern,Liu2013inplane}. Usually, when the N\'eel vector of altermagentism is perpendicular to one of the three mirror planes (along $\Gamma-\rm M_1$, $\Gamma-\rm M_2$ and $\Gamma-\rm M_3$), the mirror symmetry is preserved, ensuring the band gap being closed at the corresponding $\rm M$ point. However, here the band gap closing at three $\rm M$ points requires different altermagentism strengths. Therefore, the bulk band gap is only closed when $\phi=0, \pi$ in the first QAHE phase. The phase space is equally divided into two sections with $\mathcal{C} = 1$ and $\mathcal{C} = -1$, respectively. In the second QAHE phase, the bulk band gap closes when $\phi = n\pi/3 $ (n = 0-5). The phase space is equally divided into six sections, where the Chern number can be tuned from $\pm 1$ to $\pm 3$. 

Here, we derive an effective Hamiltonian to further understand the evolution of $\phi$-dependent Chern number in the second QAHE phase~\cite{Qiao2012}. Based on the effective Hamiltonian, we calculate the local Chern numbers around the three different $\rm M$ points as functions of azimuthal angle $\phi$ ($0<\phi<\pi$) of the N\'eel vector as shown in Fig.~S8. When $\lambda = 0.55t$, the local Chern numbers around $\rm M_1$ and $\rm M_3$ points take opposite values 1 or -1, respectively, in the intervals $\phi < \pi/3$ or $\phi > 2\pi/3$; but take same values 1 in the intervals $\pi/3<\phi < 2\pi/3$. The local Chern number around $\rm M_2$ always equals 1 when $0< \phi < \pi$. Therefore, the total Chern number $\mathcal{C} = \sum_{v=1}^3 \mathcal{C_{\rm M_v}} = 1, 3$ and $1$, respectively, in the intervals $\phi < \pi/3$, $\pi/3<\phi < 2\pi/3$ and $\phi > 2\pi/3$. These are consistent with our numerical results.
\begin{figure}[t]
	\centering
	\includegraphics[width=8 cm,angle=0]{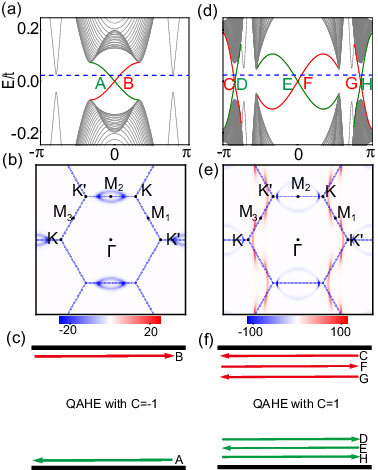}
	\caption{(a) and (d) Band structures of zigzag edge nanoribbons when the Néel vector of altermagnetism is along the $z$ direction with $\lambda/t = 0.3, 0.6$, respectively. Edge states located at opposite boundaries are highlighted by red and green curves, respectively. The other parameters are set to be $t_{\mathrm{IR}} = t_{\mathrm{SO}} = 0.1t$. (b) and (e) Corresponding Berry curvatures distributions.  (c) and (f) Schematic of edge state propagation in the zigzag edge geometry for the case of different QAHE phases (a) $\mathcal{C} = -1$ and (d) $\mathcal{C} = 1$. Labels A-H correspond to band labels in (a) and (d).}
	\label{Fig3}
\end{figure}

When the N\'eel vector of altermagnetism points to the $z$ direction, the bulk band gap also experiences closing and reopening twice for $\lambda=0.25t$, and $0.50t$ (see Fig.~S2), respectively. The process of band gap closing is identical to the case of in-plane altermagnetism. In the first QAHE region, i.e., $0.25t<\lambda<0.50t$, there is a pair of edge states in the bulk band gap with Chern number $\mathcal{C} = -1$. The corresponding Berry curvature is mainly located around $\rm M_2$ point (see Fig.~\ref{Fig3}(b)), and one pair of edge states emerges at system boundaries as displayed in Fig.~\ref{Fig3}(c).
In the second QAHE region, i.e., $\lambda>0.50t$, it is noteworthy that three pairs of edge states with mixed chirality emerge at the zigzag nanoribbon with a Chern number of $\mathcal{C} = 1$ (see Fig.~\ref{Fig3}(d)). The corresponding Berry curvatures around $\rm M_1$ and $\rm M_3$ points have opposite signs to that around $\rm M_2$ point (See Fig.~\ref{Fig3}(e)). As displayed in Fig.~\ref{Fig3}(f), the three pairs of edge states C(D), E(F), and G(H) possess different propagation directions localized at the same boundary. These three pairs of edge states are also observed in an armchair nanoribbon (see Fig.~S3). To further understand the topological phase transition with out-of-plane altermagnetism, we draw the phase diagram in the ($t_{\mathrm{IR}}$, $\lambda$) plane. Three topological phases sequentially emerge in the parameter spaces, i.e., (i) the second-order TI, (ii) the Chern insulator with $\mathcal{C} = -1$, and (iii) the Chern insulator with $\mathcal{C} = 1$ [See phase diagram in Fig.~S4(b)].

\textit{Summary}.---By introducing the $d$-wave altermagnetism, we establish a modified Kane-Mele model possessing rich topological phases. When the altermagnetism lies in $x-y$ plane, with the increase of the strength, the system first enters a second-order TI phase with corner states located at obtuse diagonals in a zigzag-edged finite-size rhombic sheet. After undergoing the first bulk band gap closing and reopening, the system enters one QAHE phase with Chern number $\mathcal{C} = \pm 1$. After further undergoing the second bulk band gap closing and reopening, the system enters the other QAHE phase where Chern number $\mathcal{C}$ equals $\pm 1$ or $\pm 3$, depending on the direction of in-plane altermagnetism. When the altermagnetism is along $z$ direction, the system also undergoes three topological phase transitions with the increase of the strength. Notably, in the unconventional QAHE phase, there are three pairs of edge states with mixed chiralities. Our work demonstrates that altermagentism provides an alternative tool to effectively tune the Kane-Mele $\mathbb{Z}_2$ TI into second-order TI and Chern number tunable QAHE phases.

\textit{Acknowledgements}.--- This work was financially supported by the National Natural Science Foundation of China (12474158, 12488101, and 111974327), Anhui Initiative in Quantum Information Technologies (AHY170000), Innovation Program for Quantum Science and Technology (2021ZD0302800) and China Postdoctoral Science Foundation (2023M733411 and 2023TQ0347). We also thank the Supercomputing Center of University of Science and Technology of China for providing high-performance computing resources.

\end{document}